\def\alphas{\alpha_{\rm s}}
\def\lstop{l_{\rm stop}}
\def\qhat{\hat q}

\documentclass[]{iopart}

\usepackage{graphicx}
\begin{document}

\title[Some new results for ``jet'' stopping in AdS/CFT]
{Some new results for ``jet'' stopping in AdS/CFT\\
 (long version)%
\footnote{Talk given at Quark Matter 2011, Annecy, France, May 23--30, 2011.
  This is a long version of the 4-page paper submitted for
  publication in the conference proceedings.
}}

\author{Peter Arnold and Diana Vaman}

\address{
    Department of Physics,
    University of Virginia, Box 400714,
    Charlottesville, Virginia 22904, USA
}
\ead{\mailto{parnold@virginia.edu}, \mailto{dv3h@virginia.edu}}
\begin{abstract}
  We give a breezy, qualitative overview of some of our recent
  results \cite{adsjet,adsjet2}
  on studying jet stopping in strongly-coupled plasmas using
  gauge-gravity duality.  Previously, people have found that the
  maximum stopping distance in such plasmas scales with energy
  as $E^{1/3}$.  We show that there is an important distinction
  between typical and maximum stopping distances.  For the
  strongly-coupled excitations that we study, we find that the
  typical stopping distance scales with energy as $E^{1/4}$.
\end{abstract}



\begin {figure}
\begin {center}
  \includegraphics[scale=0.4]{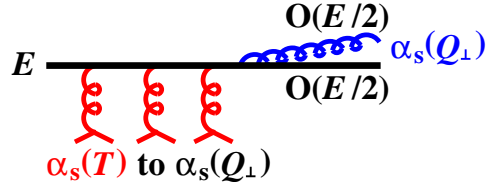}
  \caption{
     \label{fig:brem}
     Weak-coupling picture of energy loss due to bremsstrahlung.
  }
\end {center}
\end {figure}

How far does a high-energy excitation travel through a quark-gluon
plasma, before it stops in the medium and thermalizes?
To get the discussion started, consider the weak-coupling picture of
energy loss depicted in fig.\ \ref{fig:brem}, which is dominated
by nearly-collinear bremsstrahlung at high energy.
What sets the scale for the running coupling constant in this
process?  As the particle flies along, it picks up small kicks
from the medium, which change its direction, which is a form of
acceleration, and so a source for bremsstrahlung.  The individual
kicks are dominated by relatively soft momentum scales for which
$\alphas$ is of order $\alphas(T)$ (shown in red).
But now consider the $\alphas$ (shown in blue) associated with the emission of
the bremsstrahlung gluon.  Its scale is plausibly set by the
relative transverse momenta $Q_\perp$ of the two daughter particles after
the splitting.%
\footnote{
  More specifically, $Q_\perp$ is the typical transverse momentum
  picked up during the formation time of the bremsstrahlung gluon.
}
In an infinite medium (the relevant case if
the medium is thick enough for a jet to actually stop), the typical
$Q_\perp$ grows slowly with particle energy as
$Q_\perp \sim (\hat q E)^{1/4}$.%
\footnote{
  See, for example, eq.\ (4.4b) of ref.\ \cite{BDMSquench}, here parametrically
  extrapolated to the case $\omega\sim E$.  This relation
  follows from the usual LPM relations that (for the case $\omega\sim E$
  of interest) $t_{\rm form} \sim E/Q_\perp^2$
  and (for an infinite medium) $Q_\perp^2 \sim \hat q t_{\rm form}$,
  where $t_{\rm form}$ is
  the formation time.
}
The scaling of the stopping distance $\lstop$ depends on the size of
$\alphas$ at these two scales $T$ and $Q_\perp$.
\begin {itemize}
\item {\it weak coupling}:
  If $\alphas(T) \sim \alphas(Q_\perp)$ are both small, then
  $\lstop \propto E^{1/2}$ (up to logarithms).  This scaling is
  a corollary of the early work of Baier et al.\ (BDMPS) \cite{BDMPS}
  and Zakharov \cite{Zakharov} on the LPM effect in QCD.
  (For specific formulas for the stopping distance in weakly-coupled
  QCD, see ref.\ \cite{stopping}.)
\item {\it mixed coupling}:
  If $\alphas(T)$ is large and $\alphas(Q_\perp)$ is small, various
  people have argued that the stopping distance still scales as
  $\lstop \propto E^{1/2}$ and that the strong-coupling dynamics of
  the problem can be isolated into the value of $\qhat$.
  This is related, for example, to the motivation of
  Liu, Rajagopal, and Wiedemann \cite{LRW} to find the value of $\qhat$
  in strongly-coupled theories via AdS/CFT.
\item {\it all strong coupling}:
  For $\alphas(T) = \alphas(Q_\perp)$ very large,%
\footnote{
    More precisely $N_{\rm c} \alphas(T) = N_{\rm c} \alphas(Q_\perp)$
    very large for large-$N_{\rm c}$ ${\cal N}{=}4$ super Yang Mills.
}
  progress has been
  made in theories such as ${\cal N}{=}4$ super Yang Mills (and its
  close cousins), where calculations are possible using gauge-gravity
  duality.  In this case, people have found that the stopping
  distance has a different scaling with energy: $\lstop \propto E^{1/3}$
  \cite{GubserGluon,HIM,CheslerQuark}.
\end {itemize}
In this article, we will focus on the last, all strong-coupling case.
What is interesting about this result is its demonstration that the
exponent $\nu$ in $\lstop \propto E^\nu$ can depend on $\alphas$.
We will see that the strong-coupling case has an even richer
set of exponents than $\lstop \propto E^{1/3}$ to describe
jet stopping.

\begin {figure}
\begin {center}
  \includegraphics[scale=0.4]{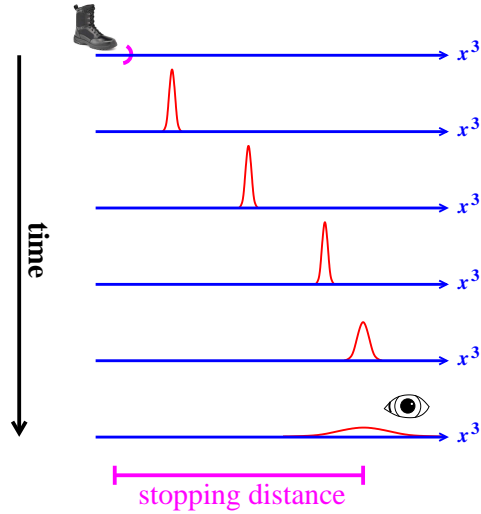}
  \caption{
     \label{fig:kick}
     Evolution in time of a local excitation of some conserved charge
     created by a high-energy kick to the plasma.  Here $x^{\textsf{3}}$ means
     the third spatial coordinate, and the eyeball represents a
     late-time observation of the system.
  }
\end {center}
\end {figure}

Think about the finite-temperature 3+1 dimensional field theory, and
imagine what happens if you very energetically
kick the medium is some way at $t{=}0$,
creating a localized excitation that moves through the medium, as
depicted in fig.\ \ref{fig:kick}.  At later times, one looks at the
progress of the excitation through the medium by following
some conserved density such as energy or
momentum or charge density.  For the purpose of this figure, it will
be simplest if we think of charge density.
As time passes, the charge density moves in a pulse that
eventually slows down (because of energy loss) and then
stops as the energy in the excitation thermalizes, after which
the charge diffuses outward.  The distance from the kick to the
center of the late-time diffusion cloud of charge is the stopping
distance.

\begin {figure}
\begin {center}
  \includegraphics[scale=0.4]{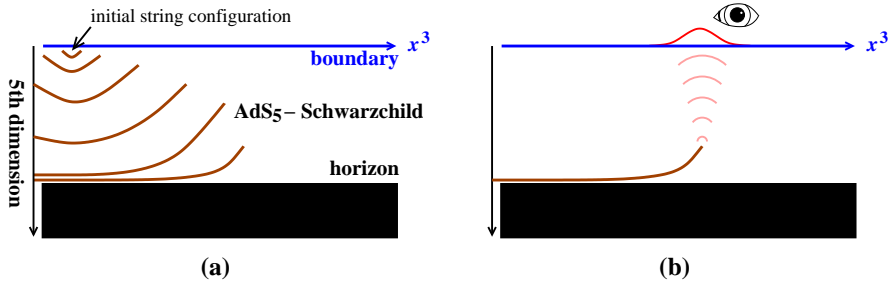}
  \caption{
     \label{fig:string}
     A cartoon of (a) the falling string of Chesler et. al \cite{CheslerQuark}
     and (b) its effect on baryon number charge density on the boundary.
     See ref.\ \cite{CheslerQuark} for their own versions of these cartoons.
  }
\end {center}
\end {figure}

As an example of previous calculations of jet stopping in AdS/CFT which
found $\lstop \propto E^{1/3}$, we briefly review in fig.\
\ref{fig:string}
how Chesler, Jensen, Karch, and Yaffe \cite{CheslerQuark} set up the
problem for light-quark jets.
A classical string in the 5-dimensional gravity
theory is used to describe a state analogous to a $q \bar q$ pair
in QCD.  The string is initialized in a way that its end points
fly apart, analogous to a high-energy back-to-back quark and
anti-quark.  As the ends fly apart, the string also falls in
the fifth dimension towards the black brane horizon, as depicted
by the time-series of curves, from top to bottom, in
fig.\ \ref{fig:string}a.  The presence of the string
disturbs the background gravitational and other fields, and so causes
a deformation of the fields on the boundary, as depicted
in fig.\ \ref{fig:string}b.
As in fig.\ \ref{fig:kick}, it will be pictorially simplest to
discuss observing conserved charge densities which diffuse
(rather than energy or momentum density, which can create sound
waves).  In the case of Chesler et.\ al, the relevant global charge
is ``baryon'' number.  As the string falls toward the horizon, its
effect on the boundary becomes more and more red-shifted, and so
spreads out to longer and longer wavelengths.  This is how the
falling string produces diffusion of the charge density seen in the
boundary theory.

There is one just slightly dissatisfying aspect of this very nice
picture: The initial configuration of the system has been specified
in the language of the gravity dual rather than directly in the
3+1 dimensional field theory.
Our goal is to avoid this by finding a way to formulate
some ``jet'' stopping
problem from beginning to end in the field
theory.  We will only use the gravity dual as a means to solve
that field-theory problem.

\begin {figure}
\begin {center}
  \includegraphics[scale=0.4]{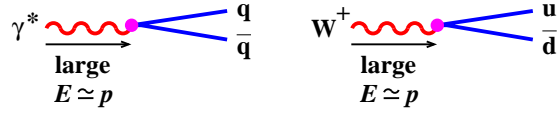}
  \caption{
     \label{fig:decay}
     The decay of a high-energy slightly-virtual photon, or of a
     high-energy $W^+$, in a QCD quark-gluon plasma.
  }
\end {center}
\end {figure}

To understand our method, it helps to first think impessionistically
of the decay of a very high energy, slightly virtual photon, or of
a very high energy $W^+$ boson, in
a QCD quark-gluon plasma.
Fig.\ \ref{fig:decay} shows the corresponding
processes for weak coupling.
Each creates a localized, high-energy, high-momentum excitation in
the plasma, moving towards the right of the figure
(the $x^\textsf{3}$ direction).  We will loosely call such an excitation
a ``jet.''  In weakly-coupled QCD, the jet produced in fig.\
\ref{fig:decay}
starts out as a nearly-collinear
$q\bar q$ pair.  In strongly-coupled ${\cal N}{=}4$ SYM, the
analogy is a localized excitation of strongly interacting gluons, gluinos,
and adjoint scalars.

Now treat the red photon or $W$ line in
fig.\ \ref{fig:decay} as an external field in the form of a high-energy
plane wave $e^{i\bar k\cdot x}$ with $k^\mu = (E,0,0,E)$.
However, we will also approximately localize that external field, so
that we know where the jet it creates starts from.
That way, when we later see where the jet stops and thermalizes,
we will be able to extract the stopping distance from the difference.
Put all together, we replace the photon or $W$ by adding a source
term to the Lagrangian:
\begin {equation}
   {\cal L}_{\rm QFT} \to
   {\cal L}_{\rm QFT}
   + {\cal O}(x) \, \Lambda_L(x) \, e^{i\bar k\cdot x}
\label {eq:source}
\end {equation}
with
\begin {equation}
   \bar k^\mu = (E,0,0,E) .
\end {equation}
The factor $\Lambda_L(x) \, e^{i\bar k\cdot x}$ is the external field,
where $\Lambda_L(x)$ is a smooth envelope function which localizes
the source to a region of size $L$ around the origin in both space
and time.  The ${\cal O}(x)$ is
the field-theory operator the external field couples to,
such as $j_\mu(x)$ in the examples of fig.\ \ref{fig:decay},
and corresponds to the vertex in fig.\ \ref{fig:decay}.

\begin {figure}
\begin {center}
  \includegraphics[scale=0.4]{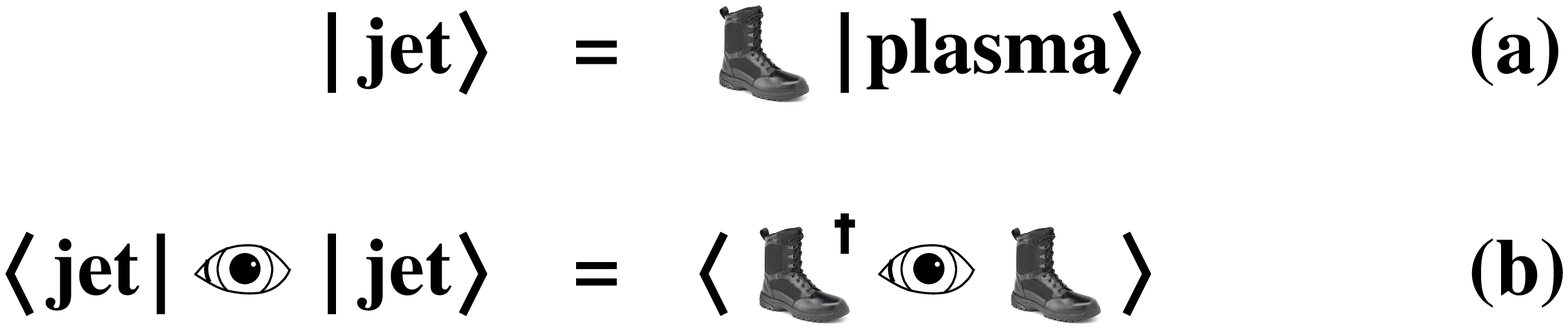}
  \caption{
     \label{fig:3point}
     Schematic picture of how the expectation of an
     observable (eyeball) in a jet state
     created by some operator (boot) can be related to a 3-point
     equilibrium correlator.
  }
\end {center}
\end {figure}

It turns out that the response to such a source is measured by
a {\it 3-point correlator}\/ in the field theory.  A crude but
simple way to understand this is given in fig.\ \ref{fig:3point}.
The boot represents the source operator in (\ref{eq:source}).
We apply this operator to a generic state of the thermal plasma
to create our localized, high-energy ``jet'' state, as in
fig.\ \ref{fig:3point}a.  We then want the expectation value in
this state of some conserved
density---energy or momentum or charge---represented by the
eyeball operator in fig.\  \ref{fig:3point}b.  Substituting
fig.\ \ref{fig:3point}a into the left-hand side of
fig.\ \ref{fig:3point}b yields the right-hand side of
fig.\ \ref{fig:3point}b, which is a 3-point correlator
of the boot, the eyeball, and the boot-conjugate, calculated
for an equilibrium plasma.%
\footnote{
   The picture presented here is intended only for the sake of
   giving the reader a flavor for why 3-point correlators are
   involved.  A better argument, in terms of response theory,
   is given in Sec.\ II.B of ref.\ \cite{adsjet2}.
   The motivation given here, however, is somewhat akin to
   that of the zero-temperature work of Hofman and Maldacena
   \cite{HofmanMaldacena}.  See Sec.\ I.C of ref.\ \cite{adsjet2}
   for a discussion of similarities and differences.
}
For {\it finite-temperature}\/ AdS/CFT
calculations, there is a great deal in the literature on computing
2-point correlators, but until recently almost nothing
carried through to specific results for 3-point correlators.

\begin {figure}
\begin {center}
  \includegraphics[scale=0.4]{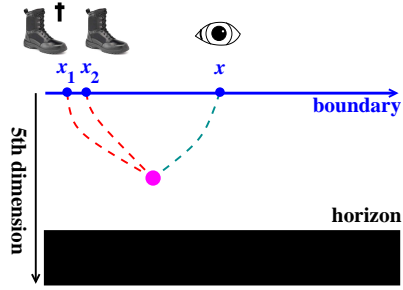}
  \caption{
     \label{fig:ADS3point}
     3-leaf tree diagram that gives the 3-point correlator in
     the gravity dual theory.
  }
\end {center}
\end {figure}

To compute this field-theory three-point correlator, we use
gauge-gravity duality to relate it to a three-point boundary
correlator in AdS$_5$-Schwarzschild space, as depicted in
fig.\ \ref{fig:ADS3point}.  Here $x$ is the space-time point where
we make our observation, and $x_1$ and $x_2$ are integrated over
the source region.  Recall that gauge-gravity duality relates
the strongly-coupled field theory problem to a problem in
{\it classical}\/ gravity, which means tree diagrams in the
gravity problem.  In Fig.\ \ref{fig:ADS3point}, we therefore
have a three-leaf tree diagram made up of the dashed lines
meeting at a vertex in the bulk.  The vertex represents a cubic
term in the 5-dimensional supergravity action.  The dashed lines
are bulk-to-boundary propagators and in our calculation are given
by {\it Heun}\/ functions.  Heun functions are obscure generalizations
of hypergeometric functions which are difficult to work with.
As a result, it is difficult to make analytic progress in
calculating generic 3-point functions, which requires integrating
over the vertex position.  It is also difficult to evaluate such
integrals numerically in our problem because the fact that our
source is high energy, and the fact that we are looking at
real-time correlations, means that the bulk-to-boundary
propagators associated with the boots are very highly oscillatory
functions and so very difficult to integrate.

Fortunately, this very problem also points the way to a solution.
The (red) dashed lines in fig.\ \ref{fig:ADS3point} associated
with the boots are high-energy/momentum propagators, and we
may make a corresponding high-$k$ WKB-like approximation, which
gives them a relatively simple analytic form.
For the remaining (green) dashed line in fig.\ \ref{fig:ADS3point},
recall that
the technique for measuring the stopping distance is to
look for the spatial center of late-time charge diffusion,
as depicted by the last
curve in Fig.\ \ref{fig:kick}.  This corresponds to
measurements of large wavelengths and so small wave numbers
of the late-time response of the system.  As a result, we may
use a low-$k$ approximation to the (green) dashed line
associated with the eyeball in Fig.\ \ref{fig:ADS3point}.
The low-$k$ limit also gives a relatively simple
analytic expression for the propagator.
These approximations,
appropriate for the problem we want to solve, turn out
to be enough to make it possible to analytically integrate
over the bulk vertex position and compute the 3-point correlator.
[By the way, it is easy to see now why this physics could not
possibly be captured by a 2-point correlator.  If we took a
2-point correlator of a boot (source) and an eyeball (measurement),
there would be
a mismatch between the high-momentum of the boot and the
low wave-numbers of interest for the eyeball, and so the
2-point correlator would vanish by momentum conservation.]

\begin {figure}
\begin {center}
  \includegraphics[scale=0.4]{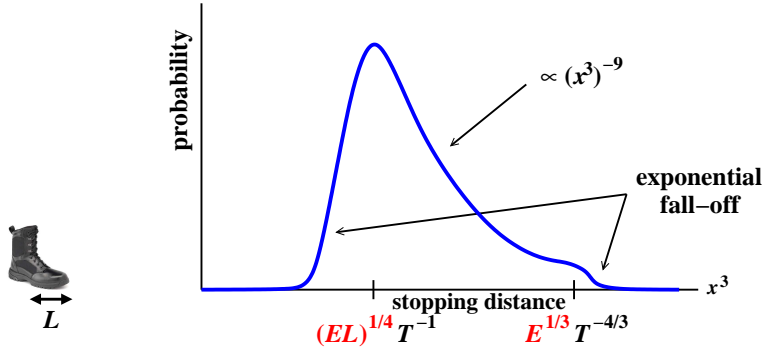}
  \caption{
     \label{fig:result}
     Qualitative picture of the probability distribution of
     jet stopping distances, where $L$ is the source size.
     The power-law fall-off $(x^{\textsf{3}})^{-9}$ shown
     for the intermediate region is for the specific source
     operator analyzed in ref.\ \cite{adsjet}; the more general
     case is given in ref.\ \cite{adsjet2}.
  }
\end {center}
\end {figure}

The results \cite{adsjet,adsjet2} of using this method to
calculate the jet stopping distance are summarized in Fig.\
\ref{fig:result}, which shows a probability distribution for
the jet stopping distance.  There is indeed a ``maximum'' stopping
distance that scales like $E^{1/3}$, beyond which the probability
distribution falls off exponentially.  However, this is not
the {\it typical}\/ stopping distance of the jets created.
Almost all of the jets created by our kick (\ref{eq:source})
instead stop sooner at a distance scale proportional to
$(EL)^{1/4}$, where $L$ is the size of the space-time region
in which the jet was initially created (i.e. the size of the toe
of our boot and the duration of the kick).%
\footnote{
  In the subject of jet quenching, the letter $L$ is often instead used to
  stand for the thickness of the medium.  This has nothing to
  do with our $L$ here: in our problem, we are
  considering jet stopping in an infinite medium.
}
Between the scales $(EL)^{1/4}$ and $E^{1/3}$, the probability
distribution falls algebraically, with the power law depending
on the conformal dimension of the source operator \cite{adsjet2}.

By the way, for the $x^3 \ll E^{1/3} T^{-4/3}$ case, which
dominates the probability distribution shown in fig.\ \ref{fig:result},
the calculation can be made even simpler than
the forgoing discussion.  In this limit, it turns out that
one can use the approximation of geometric optics and replace
the 3-point calculation of fig.\ \ref{fig:ADS3point} by a
simple classical calculation of a
5-dimensional particle falling toward the horizon
in AdS$_5$-Schwarzschild space.  See ref.\ \cite{adsjet2}
for details.

\begin {figure}
\begin {center}
  \includegraphics[scale=0.4]{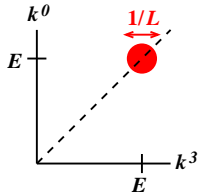}
  \caption{
     \label{fig:momentum}
     Qualitative picture of the region in 4-momentum where the
     source (\ref{eq:source}) has support.
  }
\end {center}
\end {figure}

Readers may wonder what the size $L$ of the source has to do with
how far jets propagate?  The answer is that $L$ determines the
virtuality of the source.  Recall that in (\ref{eq:source}) we
chose a source proportional to $\Lambda_L(x) \, e^{i\bar k\cdot x}$
with $\bar k^\mu = (E,0,0,E)$.  The envelope function $\Lambda_L(x)$
localized the source to a region of size $L$ and so, by the
uncertainty principle, introduces a spread in momentum of size
$1/L$.  Fig.\ \ref{fig:momentum} gives a qualitative picture
of the region of momentum space that contributes to
$\Lambda_L(x) \, e^{i\bar k\cdot x}$.  Generic momenta
in this region are not exactly on the light cone, and we
demonstrate in ref.\ \cite{adsjet2} that the $(EL)^{1/4}$
scaling of the typical stopping distance really means
$(E^2/q^2)^{1/4}$, where $q^2$ is the typical virtuality of
the source:
\begin {equation}
   l_{\rm typical} \sim \left( \frac{E^2}{q^2} \right)^{1/4} \frac{1}{T}
   \,.
\label {eq:stop}
\end {equation}

Studying jet stopping in the strongly-coupled limit is a useful
theoretical exercise in order to sharpen our understanding of
the role that coupling $\alphas$ plays in the scaling of jet
stopping with energy.
It also provides an interesting test case for optimists, or a
straw man for cynics, for phenomenological comparisons with data:
Regardless of one's point of view, AdS/CFT results inspire
the question of whether
experiment can differentiate between $E^{1/2}$ and
$E^{1/3}$ or other scaling for jet stopping distances, or
between non-standard scaling laws for other measurements related to
energy loss.
For some work along these lines, see refs.\ \cite{jet1,jet2,jet3}.
Readers who toy with such questions might naturally ask
what happens if we now put $q^2{\sim}E^2$
into (\ref{eq:stop}), since $q^2{\sim}E^2$ is
the virtuality scale of processes creating high-energy jets
at mid-rapidity in actual relativistic heavy ion collisions.  The
corresponding running coupling constant associated with the initial
moments of such jet creation is $\alphas(E)$, rather than
the couplings $\alphas(T)$ or $\alphas(Q_\perp)$ reviewed in the introduction.
Whatever one may think of the theoretical or perhaps phenomenological
utility of investigating jet stopping in strongly-coupled theories,
it seems implausible that $N_{\rm c} \alphas(E)$
can be approximated as infinite for high
energy jets.  If there is any phenomenological application for the
results presented here, one should likely avoid treating the system as strongly
coupled until enough time has passed that the initial
virtuality $q^2{\sim}E^2$ associated with jet creation
has dropped to some lower virtuality where one is willing to at least
entertain the possibility that the
$\alphas$ associated with the emission of a bremsstrahlung
gluon might be effectively strongly coupled.


\ack
This work was supported, in part, by the U.S. Department
of Energy under Grant No.~DE-FG02-97ER41027 and by a
Jeffress research grant, GF12334.


\end{document}